\documentclass[a4paper,onecolumn,11pt,accepted=2026-08-02]{quantumarticle}
\pdfoutput=1

\usepackage[utf8]{inputenc}
\usepackage[english]{babel}
\usepackage[T1]{fontenc}
\usepackage{amsmath,amssymb}
\usepackage{hyperref}

\usepackage{tikz}
\usetikzlibrary{quantikz2}

\usetikzlibrary{shapes,decorations.pathreplacing,calligraphy}

\usepackage{xcolor}

\title{Catalytic $z$-rotations in constant $T$-depth}
\author{Isaac H. Kim}
\affiliation{Department of Computer Science, University of California, Davis, CA, 95616, USA}

\begin{document}

\maketitle

\begin{abstract}
 We show that the $T$-depth of any single-qubit $z$-rotation can be reduced to $3$ if a certain catalyst state is available. To achieve an $\epsilon$-approximation, it suffices to have a catalyst state of size polynomial in $\log(1/\epsilon)$.  This implies that $\mathsf{QNC}^0_f/\mathsf{qpoly}$ admits a finite universal gate set consisting of Clifford+$T$. In particular, there are catalytic constant $T$-depth circuits that approximate multi-qubit Toffoli, adder, and quantum Fourier transform arbitrarily well. We also show that the catalyst state can be prepared in time polynomial in  $\log (1/\epsilon)$.
\end{abstract}

\section{Introduction}

A fundamental result in quantum computing is the Solovay-Kitaev theorem~\cite{kitaev2002classical}, which states that a finite set of universal gates can approximate the continuous gate set arbitrarily well, with an overhead that scales polylogarithmically with the inverse precision. While the Solovay-Kitaev theorem applies to any universal gate set, there are more efficient methods tailored to specific gate sets. A common example in the literature is Clifford+$T$, or alternatively, Clifford+Toffoli. Both gate sets can be realized in a fault-tolerant manner using a variety of methods, making them the standard gate set for fault-tolerant quantum computation. 

For a long time, a common cost metric used for these gate sets has been the number of non-Clifford gates. This is because the cost of these gates turned out to be substantially more expensive than that of the Clifford gates; see Ref.~\cite{litinski2019magic} and the references therein. In particular, the $T$-count or Toffoli-count has been the primary metric to minimize. From this perspective, there are essentially optimal methods for approximating single-qubit $z$-rotations~\cite{kliuchnikov2013asymptotically,ross2016optimalancillafreecliffordtapproximation,bocharov2015efficient,kliuchnikov2023shorter}; they all possess the optimal $\mathcal{O}(\log (1/\epsilon))$ $T$-count (or equivalently, Toffoli-count), where $\epsilon$ is the approximation error. 

However, if the goal is to minimize the computation time, one should instead optimize the $T$-depth. In the surface code-based approach to fault-tolerant quantum computation~\cite{Fowler2012}, there is a standard space-time tradeoff one can make~\cite{fowler2013timeoptimalquantumcomputation,Litinski2019gameofsurfacecodes}. Keeping the overall space-time volume of the computation as a constant, it is possible to reduce time by using more space. Once the time overhead is reduced to the maximum extent possible, the computation time is essentially limited by the $T$-depth of the underlying circuit. Therefore, the $T$-depth of a circuit determines the minimal time needed to execute the circuit on a fault-tolerant quantum computer. Recent works also show that the cost of $T$-gates can be much cheaper than previously anticipated, suggesting that this metric may become practically relevant in the future~\cite{goto2016minimizing,Chamberland2020,Itogawa2024,gidney2024magicstatecultivationgrowing,Daguerre2025}; see Ref.~\cite{daguerre2025experimentaldemonstrationhighfidelitylogical,dasu2025breakingmagicdemonstrationhighfidelity} for the recent experiments.

Thus it is natural to ask what the achievable time overhead of single-qubit rotation is, quantified in terms of the $T$-depth. To that end, it is worth noting that the number of $T$-gates necessary to achieve $\epsilon$-approximation is $\Omega(\log(1/\epsilon))$~\cite{beverland2020lower}. While the existing methods~\cite{kliuchnikov2013asymptotically,ross2016optimalancillafreecliffordtapproximation,bocharov2015efficient,kliuchnikov2023shorter} match this lower bound, they are all sequential in nature. Is it possible to parallelize them so that the $T$-depth can be brought down to a constant? This question was raised recently in Ref.~\cite{parham2025quantumcircuitlowerbounds}.

The main purpose of this paper is to investigate this question and provide a partial but positive answer. Before we proceed further, let us make a few remarks to explain our main result clearly. For any Clifford+$T$ circuit, one can apply all the $T$-gates in one step by preparing the associated magic states and carrying out the computation by gate teleportation~\cite{Gottesman1999}. However, gate teleportation requires an adaptive basis change depending on the measurement outcome, which cannot be parallelized in general. For this reason, we focus on circuits with constant $T$-depth without any adaptive basis change. We also note that there are several different versions of this problem one can consider, depending on the types of ancilla we use. The cheapest type of ancilla is the dirty ancilla. These are qubits in an unknown state, which must be returned to the original state after the computation. The second is a clean ancilla, which is initialized to a known simple state, such as $|0\ldots 0\rangle$. The third would be a \emph{catalyst state}~\cite{Campbell2011,Gidney2019efficientmagicstate,beverland2020lower,amy2023catalyticembeddingsquantumcircuits}. This is a special state prepared offline before the computation, which is then used in the computation. These states are returned to the original state after the computation, and therefore can be reused later. 

In this paper, we show that constant $T$-depth rotation is possible using an appropriate catalyst state, answering the third (and the weakest) version of the problem in the affirmative. Specifically, we show that for every $a\in [2^n-1]$, there is a unitary $U_a$ with the following properties. First, it applies a single-qubit $z$-rotation proportional to $a$:
\begin{equation}
    U_a((\alpha |0\rangle + \beta|1\rangle)|\phi_a\rangle) = (\alpha|0\rangle + \beta e^{\frac{2\pi ai }{2^n-1}}|1\rangle)|\phi_a\rangle, \label{eq:rot_feedback}
\end{equation}
where $|\phi_a\rangle$ is a state over $\mathcal{O}(n)$ qubits. Second, $U_a$ has $T$-depth of $3$ and a Clifford depth of $\mathcal{O}(\log n)$. Third, $|\phi_a\rangle$ can be prepared in time polynomial in $n$. Therefore, the catalyst state $|\phi_a\rangle$ can be prepared with a modest cost, after which the rotation can be realized in constant $T$-depth. We also discuss a variant of this construction in which the rotation angle is not fixed to a constant. The $T$-depth remains the same for this variant, though we seem to need $n$ copies of the same catalyst state. 

Because single-qubit rotation is a widely used subroutine in quantum computing, our protocol can be used to lower the time overhead of a large class of quantum subroutines. One simple way to understand this consequence is in terms of the complexity class $\mathsf{QNC}^0_f$~\cite{hoyer2005quantum,takahashi2016collapse}. The circuits in this class are constant-depth quantum circuits assisted by unbounded fanout (which is a Clifford), using arbitrary one- and two-qubit gates. Because our result lets us simulate any one- and two-qubit gate in constant $T$-depth using a catalyst state, any circuit in $\mathsf{QNC}^0_f$ can be simulated by a constant $T$-depth circuit using a catalyst state. Interestingly, $\mathsf{QNC}^0_f$ includes a rich family of circuits. It includes unbounded fan-in, addition (which may or may not be modular), quantum Fourier transform, and even the quantum part of the factoring algorithm~\cite{hoyer2005quantum,takahashi2016collapse}. Our result implies that these can all be done in constant $T$-depth, insofar as enough catalyst states are available. 

The fact that unbounded fan-in and addition can be realized in constant $T$-depth (albeit with a catalyst) is somewhat counterintuitive. Because $T$-gates and Toffolis can be converted to each other up to a Clifford and a catalyst state, it follows that there are unbounded fan-in and adders that can be realized in constant Toffoli depth. This is in contrast to the classical case, in which the Toffoli depth of these circuits is known to be $\Omega(\log n)$~\cite{beigel1993polynomial,hoyer2005quantum}. Because Clifford gates cannot generate additional Toffolis, it seems somewhat surprising that they can help us reduce the Toffoli depth of these circuits. 

So, what makes this counterintuitive result possible? A broad direction we take in our construction is to use a specialized form of in-place adder. To provide more context, let us recall the well-known method for implementing single-qubit rotation with an adder~\cite{kitaev2002classical,gidney_halving_2018}. Let $\mathsf{ADD}_a$ be a unitary that implements $\mathsf{ADD}_a|x\rangle = |(x+a) \mod 2^n\rangle$. Because $|QFT\rangle:= \frac{1}{\sqrt{2^n}}\sum_{j=0}^{2^n-1} e^{\frac{-2\pi i j}{2^n}} |j\rangle$ is an eigenstate of $\mathsf{ADD}_a$ with an eigenvalue of $e^{\frac{2\pi i a}{2^n}}$, applying the phase kickback trick to $\mathsf{ADD}_a$ and $|QFT\rangle$, we can implement the single-qubit $z$-rotation with an angle of $\frac{2\pi a}{2^n}$. Because there is a logarithmic-depth in-place adder~\cite{gidney_halving_2018}, the phase kickback unitary can be implemented in logarithmic depth. This observation lets us implement single-qubit $z$-rotation using $\mathcal{O}(\log (1/\epsilon))$ qubits in depth $\mathcal{O}(\log \log (1/\epsilon))$~\cite[Theorem 13.5]{kitaev2002classical}.

The main difficulty is that there is no known in-place reversible adder that can be implemented in constant depth, even if one deviates from the standard binary representation. For instance, there is a constant-depth adder using signed-digit representation~\cite{avizienis2009signed}, but it is unclear if this method can be made both reversible and in-place. A new idea seems necessary to break the $\mathcal{O}(\log \log (1/\epsilon))$ barrier.

The main idea that lets us overcome this barrier is to use special polynomials over $\mathbb{F}_2$, the \emph{primitive polynomials} of $\mathrm{GF}(2^n)$. It turns out that there is a simple recurrence relation inherited from this polynomial which simultaneously satisfies a number of nice properties: (i) the recurrence relation has a period of $2^n-1$, (ii) it has a Clifford implementation, and (iii) it has a constant-depth implementation. In effect, we have an $n$-qubit Clifford that has a period of $2^n-1$ which can be implemented in constant depth. By preparing an eigenstate of this Clifford and applying phase kickback, we obtain the desired operation. The fact that this Clifford is constant-depth guarantees us that the phase kickback operation can be implemented in constant $T$-depth.

The rest of the paper is structured as follows. In Section~\ref{sec:primitive_polynomial}, we explain our constant $T$-depth protocol for fixed single-qubit rotation. We provide a discussion and an outlook in Section~\ref{sec:discussion}.

\section{Primitive Polynomial Method}
\label{sec:primitive_polynomial}
Recall that $\mathrm{GF}(2^n) = \mathbb{F}_2[x]/\langle f(x)\rangle$, where $f(x)$ is a monic irreducible polynomial of degree $n$. The elements of $\mathrm{GF}(2^n)$ can be conveniently written in terms of the root of $f(x)$, say $\alpha$.
\begin{equation}
    \mathrm{GF}(2^n) = \left\{ \sum_{j=0}^{n-1} c_j\alpha^j : c_j \in \mathbb{F}_2 \right\}. \label{eq:gf2n_def}
\end{equation}
We say $f(x)$ is \emph{primitive} if $\alpha$ generates the multiplicative subgroup of $\mathrm{GF}(2^n)$, consisting of the nonzero elements. Unless specified otherwise, we assume $f(x)$ is primitive throughout this paper.

Consider a recurrence relation $g_{n+1}(\alpha) = \alpha g_n(\alpha)$, where $g_{n}(\alpha) \in \mathrm{GF}(2^n)$. Because $f(x)$ is primitive, multiplication by $\alpha$ acts transitively on the set of nonzero elements. In particular, the period of the recurrence relation is $2^n-1$.

This recurrence relation induces a linear transformation on the coefficients of $g_n(\alpha)= \sum_{j=0}^{n-1} g_n^j \alpha^j \in \mathrm{GF}(2^n)$. Without loss of generality, let $f(x) = x^n + \sum_{j=0}^{n-1} f_j x^j$. (Here $f_0=1$ since otherwise $f(x)$ is not irreducible.) The recurrence relation can be written as follows:
\begin{equation}
    \begin{pmatrix}
        g_{n+1}^0 \\
        g_{n+1}^1 \\
        \vdots \\
        g_{n+1}^{n-2}\\
        g_{n+1}^{n-1}
    \end{pmatrix}
    =
    \begin{pmatrix}
        0 & 0 &   \cdots & 0 & 0 & f_0 \\
        1 & 0 &  \cdots & 0 & 0 & f_1\\
        \vdots  & \vdots & \ddots & \vdots & \vdots & \vdots \\
        0 & 0 &  \cdots & 1 & 0 & f_{n-2}\\
        0 & 0 &  \cdots & 0 & 1 & f_{n-1}
    \end{pmatrix}
    \begin{pmatrix}
        g_{n}^0 \\
        g_{n}^1 \\
        \vdots \\
        g_{n}^{n-2}\\
        g_{n}^{n-1}
    \end{pmatrix}.
\end{equation}
We denote the $n\times n$ matrix as $C_f$, which is the \emph{companion matrix} of $f(x)$. This matrix is invertible and its order is $2^n-1$. Furthermore, it can be decomposed into a product of the cyclic permutation and an upper triangular matrix:
\begin{equation}
    C_f = 
    \begin{pmatrix}
        0 & 0 &   \cdots & 0 & 0 & 1 \\
        1 & 0 &  \cdots & 0 & 0 & 0\\
        \vdots  & \vdots & \ddots & \vdots & \vdots & \vdots \\
        0 & 0 &  \cdots & 1 & 0 & 0\\
        0 & 0 &  \cdots & 0 & 1 & 0
    \end{pmatrix}
    \begin{pmatrix}
        1 & 0 &   \cdots & 0 & 0 & f_1 \\
        0 & 1 &  \cdots & 0 & 0 & f_2\\
        \vdots  & \vdots & \ddots & \vdots & \vdots & \vdots \\
        0 & 0 &  \cdots & 0 & 1 & f_{n-1}\\
        0 & 0 &  \cdots & 0 & 0 & f_{0}
    \end{pmatrix}
    \label{eq:cf_decomposition}
\end{equation}

To summarize, there are two different ways of representing elements of $\mathrm{GF}(2^n)$: (i) viewing them as a polynomial [Eq.~\eqref{eq:gf2n_def}] and (ii) as a binary vector. In Section~\ref{subsec:clifford} and~\ref{subsec:kickback}, we will take the second point of view. In Section~\ref{subsec:catalyst_prep}, the first point of view will prove to be more useful.

\subsection{Clifford Interpretation}
\label{subsec:clifford}
Because $C_f$ is a binary invertible matrix, we can define a Clifford unitary associated to it. Define $U_f$ as 
\begin{equation}
    U_f|z\rangle = |C_f z\rangle,
\end{equation}
where $z$ is an $n$-dimensional binary vector. Thanks to Eq.~\eqref{eq:cf_decomposition}, $U_f$ can be decomposed into two simple Cliffords: (i) cyclic shift $\mathsf{S}$ and (ii) a CNOT gate (denoted as $\mathsf{CX}$) with multiple targets. More specifically, let $Q_f = \{j: f_{j+1}=1, j\neq 0 \}$. The control and the targets of $\mathsf{CX}$ are $n-1$ and $Q_f$, respectively. Thus we denote the $\mathsf{CX}$ part of the circuit as $\mathsf{CX}_{n-1\to Q_f}$.

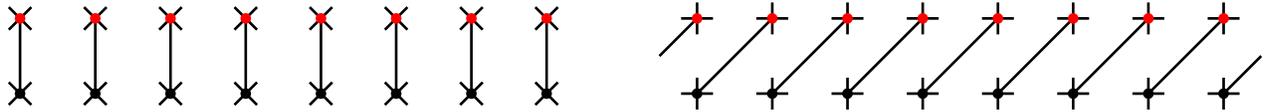
\begin{figure}[t]
    \centering
    \begin{tikzpicture}[line width=1pt, scale=0.85]
    \foreach \x in {0, ..., 7}
    {
    \draw[] (\x,0) -- (\x, 1);
    \draw[] (\x-0.15, -0.15) -- ++ (0.3, 0.3);
    \draw[] (\x-0.15, 0.15) -- ++ (0.3, -0.3);
    \draw[] (\x-0.15, 1-0.15) -- ++ (0.3, 0.3);
    \draw[] (\x-0.15, 1+0.15) -- ++ (0.3, -0.3);
    \draw[fill=black] (\x, 0) circle (1.5pt);
    \draw[fill=red, draw=red] (\x, 1) circle (1.5pt);
    }
    \begin{scope}[xshift=9cm]
    \foreach \x in {0, ..., 6}
    {
    \draw[] (\x,0) -- (\x+1, 1);
    }
    \draw[] (7, 0) -- ++ (0.5, 0.5);
    \draw[] (0, 1) -- ++ (-0.5, -0.5);
    \foreach \x in {0, ..., 7}
    {
    \draw[] (\x, -0.212) -- ++ (0, 0.424);
    \draw[] (\x-0.212, 0) -- ++ (0.424, 0);
    \draw[] (\x, 1-0.212) -- ++ (0, 0.424);
    \draw[] (\x-0.212, 1) -- ++ (0.424, 0);
    \draw[fill=black] (\x, 0) circle (1.5pt);
    \draw[fill=red, draw=red] (\x, 1) circle (1.5pt);
    }
    \end{scope}

    \end{tikzpicture}
    \caption{A depth-$2$ circuit consisting of $\mathsf{SWAP}$ gates for implementing $\mathsf{S}$. The red dots are ancillary qubits. In the right figure, a periodic boundary condition in the $x$-direction is used.}
    \label{fig:shift_highlevel}
\end{figure}

Let us discuss the decomposition of $\mathsf{S}$ and $\mathsf{CX}_{n-1\to Q_f}$ into one- and two-qubit Clifford gates. For $\mathsf{S}$, a sequential $n-1$ application of $\mathsf{SWAP}$ suffices, which yields the depth of $n-1$. Intuitively, if we have $n$ ancillary qubits prepared in an identical state, e.g., $|0\ldots 0\rangle$, it is easy to see that a depth-$2$ circuit suffices; see Fig.~\ref{fig:shift_highlevel}. Less obvious is the fact that the same depth can be achieved without using any ancillas. Define $\mathsf{Rev}_{a, b}$ as a map that reverses the order of qubits $a, ..., b$. It is straightforward to verify that
\begin{equation}
    \mathsf{S} = \mathsf{Rev}_{0, n-1}\cdot \mathsf{Rev}_{1, n-1}.\label{eq:rev}
\end{equation}
Since $\mathsf{Rev}_{a, b}$ is a depth-$1$ $\mathsf{SWAP}$ circuit, $\mathsf{S}$ can be realized by a depth-$2$ $\mathsf{SWAP}$ circuit. Moreover, the number of $\mathsf{SWAP}$s is still $n-1$.

For $\mathsf{CX}_{n-1\to Q_f}$, the decomposition $\mathsf{CX}_{n-1\to Q_f}=\prod_{j\in Q_f}\mathsf{CX}_{n-1\to j}$ yields a depth-$|Q_f|$ circuit. Because the controls of the $\mathsf{CX}$ gates overlap, it is not possible to convert this into a constant-depth circuit. However, a constant-depth implementation is possible if unbounded fanout is available. Using unbounded fanout, we can apply the following transformation:
\begin{equation}
    \alpha|0\rangle + \beta|1\rangle \mapsto \alpha |0\ldots 0\rangle + \beta|1\ldots 1\rangle
    \label{eq:fanout}
\end{equation}
using $|Q_f|$ additional ancillary qubits. Then we can apply a depth-$1$ circuit comprised of $\mathsf{CX}$ from one of the qubits in this $|Q_f|$-qubit register to a qubit in $Q_f$. Applying the inverse of the fanout, we can implement $\mathsf{CX}_{n-1\to Q_f}$. 

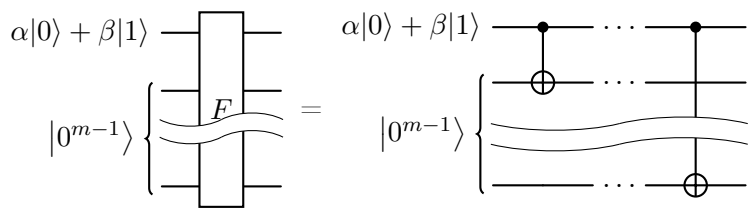
\begin{figure}[h]
    \centering
    \begin{quantikz}
        \lstick{$\alpha|0\rangle + \beta|1\rangle$} & \gate[4]{F} &  \\
        \lstick[3]{\ket{0^{m-1}}}&& \\
        \wave && \\
        &&
    \end{quantikz}
    =
    \begin{quantikz}
        \lstick{$\alpha|0\rangle + \beta|1\rangle$} & \ctrl{1} & \ \ldots \ &  \ctrl{3} &\\
        \lstick[3]{\ket{0^{m-1}}}& \targ{} & \ \ldots \ & &\\
        \wave &&  \ \ldots \ & & &\\
        && \ \ldots \ & \targ{} &
    \end{quantikz}
    \caption{Fanout gate can be realized by a sequence of $\mathsf{CX}$ gates.}
    \label{fig:fanout}
\end{figure}

Because unbounded fanout is essentially a composition of $\mathsf{CX}$ gates [Fig.~\ref{fig:fanout}], one might wonder if there is anything to be gained from it. The answer turns out to be yes when we later aim to apply a controlled version of $U_f$. The key point is that the unbounded fanout can remain as Clifford, even when the other parts of the circuit are changed to non-Clifford gates. This is crucial to achieving constant $T$-depth; see Section~\ref{subsec:kickback} for more details.

An important property of $U_f$ is that its eigenvalues nearly uniformly cover the unit circle. Because $C_f$ acts trivially on the zero element, $|0\ldots 0\rangle$ is an eigenvector of $C_f$ with an eigenvalue of $1$. The remaining eigenvectors are
\begin{equation}
    |\psi_k\rangle = \frac{1}{\sqrt{2^n-1}} \sum_{j=0}^{2^n-2} \tilde{\omega}^{-jk} |C_f^j v\rangle,
    \label{eq:catalysts}
\end{equation}
for $k\in [2^n-1]$, where $\tilde{\omega} = e^{\frac{2\pi i}{2^n-1}}$ and $v$ is a distinguished nonzero binary vector, say $v=1\ldots 0$. The eigenvalue associated to $|\psi_k\rangle$ is $\tilde{\omega}^k$.

\subsection{Phase Kickback}
\label{subsec:kickback}

Given a state $|\psi_k\rangle$, by applying the phase kickback to $U_f$, we can implement a $z$-rotation with an angle of $\frac{2\pi k}{2^n-1}$. If the goal is to apply this specific phase, then it suffices to have only $|\psi_k\rangle$. However, in many cases one would be interested in approximating an \emph{arbitrary} rotation. For that, it suffices to have $|\psi_{2^t}\rangle$ for every $t\in [n]$. Selectively applying phase kickback to these states, one can implement any single-qubit $z$-rotation with an angle of $2m\pi/(2^n-1)$, where $m\in \mathbb{Z}$.

More generally, it suffices to have $|\psi_{a2^t}\rangle$, $t\in [n]$, for any fixed $a$ that is coprime to $2^n-1$. Selectively applying phase kickback to these states, we obtain a single-qubit $z$-rotation with an angle of $2ma \pi / (2^n-1)$, where $m\in \mathbb{Z}$. Since $a$ is coprime to $2^n-1$, for any $b\in [2^n-1]$, there is some $m$ such that $ma = b \mod (2^n-1)$.

In fact, these phase kickbacks can be parallelized using unbounded fanout. After applying the unbounded fanout [Eq.~\eqref{eq:fanout}] to $n-1$ ancillary qubits, we can apply the $n$ controlled $U_f$s in parallel, followed by the inverse of the unbounded fanout. The requisite size of the catalyst state for this method is $n^2$, and if we were to parallelize the phase kickbacks, we need $n-1$ additional ancillary qubits.

\begin{figure}[h]
    \centering
    \begin{quantikz}
        \lstick{$\alpha|0\rangle +\beta |1\rangle$} & \ctrl{1} &  \\
       \lstick{$n-1$} & \ctrl{3} & \\
         \lstick[3]{$Q_f$}& \targ{} & \\
        \wave&&  \\
        & \targ{} &
    \end{quantikz}
    =
    \begin{quantikz}
        \lstick{$\alpha \ket{0} + \beta \ket{1}$} & \gate[4]{F} & \ctrl{4} & &  \ \ldots \ & & \gate[4]{F^{\dagger}} &\\
        \lstick[3]{\ket{0^{|Q_f|}}} & & & \ctrl{4} & \ \ldots \ & & & \rstick[3]{\ket{0^{|Q_f|}}}\\[0.5cm]
        \wave & & & & & & &\\
        & & & & \ \ldots \ & \ctrl{4} & &\\
        \lstick{$n-1$} & \gate[4]{F} & \ctrl{4} & & \ \ldots \ & & \gate[4]{F^{\dagger}} &\\
        \lstick[3]{\ket{0^{|Q_f|}}} & & & \ctrl{4} & \ \ldots \ & & & \rstick[3]{\ket{0^{|Q_f|}}}\\[0.5cm]
        \wave & & & & & & &\\
        & & & & \ \ldots \ & \ctrl{3} & &\\
        \lstick[3]{$Q_f$} & &\targ{} & & \ \ldots \ & & &\\
        \wave & & & & & & &\\
        & & & & \ \ldots \ & \targ{} & &
    \end{quantikz}
    \caption{Using unbounded fanout, controlled $\mathsf{CX}_{n-1 \to Q_f}$ can be implemented by a circuit with a Toffoli depth of $1$.}
    \label{fig:circ_identity}
\end{figure}
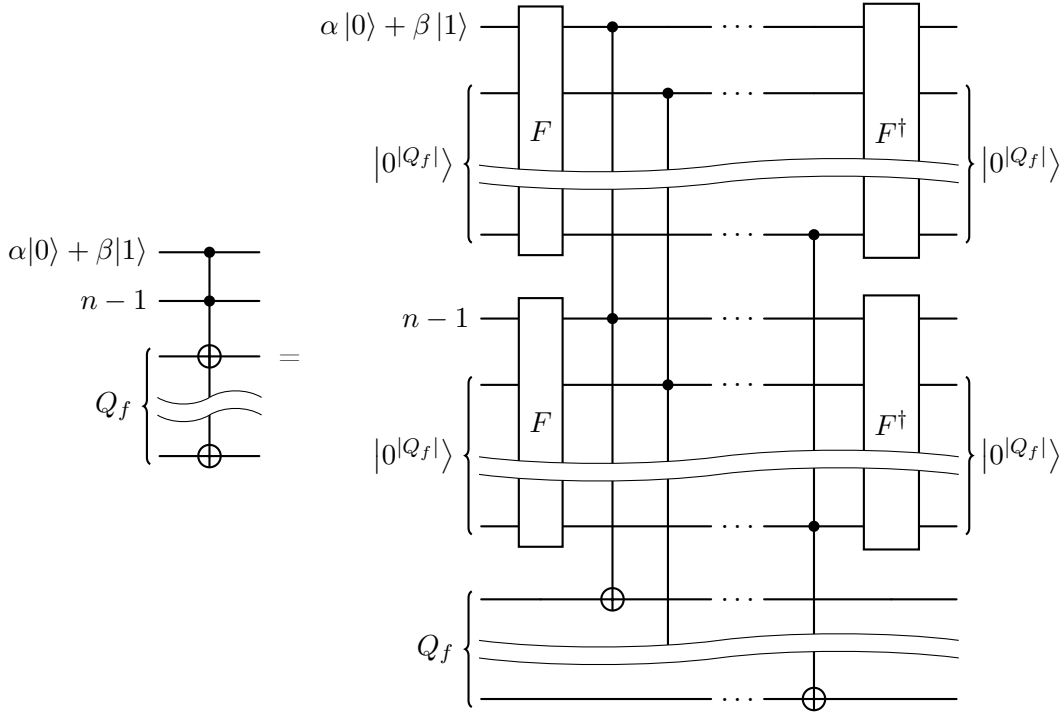

Now the remaining question is whether the controlled $U_f = \mathsf{S} \cdot \mathsf{CX}_{n-1\to Q_f}$ can be parallelized. To that end, we can again use unbounded fanout so that the control qubit is expanded to $n$ qubits. The controlled $\mathsf{S}$ operation can be realized by replacing the individual $\mathsf{SWAP}$ with controlled $\mathsf{SWAP}$, also known as the Fredkin gate. This yields a depth-$2$ circuit consisting of Fredkin gates; see Fig.~\ref{fig:shift} (top) for an example. (The same approach can be applied to Eq.~\eqref{eq:rev}.) The Fredkin gate is equivalent to a Toffoli up to Clifford [Fig.~\ref{fig:shift}, bottom], whose $T$-depth is $1$~\cite{Selinger2013}. Therefore, the controlled $\mathsf{S}$ has a $T$-depth of $2$. 
For $\mathsf{CX}_{n-1\to Q_f}$, the fanout can remain as is. We only need to change the $\mathsf{CX}$ gates to $\mathsf{CCX}$, which is Toffoli. The circuit identity is shown in Fig.~\ref{fig:circ_identity}. Thus the $T$-depth of $\mathsf{CX}_{n-1\to Q_f}$ is $1$. Altogether, the $T$-depth of the controlled $U_f$ is $3$. 

\begin{figure}[t]
    \centering
    \begin{quantikz}
\lstick[1]{$\alpha \ket{0} + \beta \ket{1}$} &\gate[5]{F} & \ctrl{5}\gategroup[15, steps=5,style={dashed,rounded
    corners,fill=blue!20, inner
    xsep=2pt},background,label style={label
    position=below,anchor=north,yshift=-0.2cm}]{{First layer}}  & &\ \ldots\ & & & \ctrl{5}\gategroup[15, steps=5,style={dashed,rounded
    corners,fill=red!20, inner
    xsep=2pt},background,label style={label
    position=below,anchor=north,yshift=-0.2cm}]{{Second layer}} & & \ \ldots \  & & & \gate[5]{F^{\dagger}} &\\  
\lstick[4]{\ket{0^{n-1}}}&&& \ctrl{5}& \ \ldots \ &&& & \ctrl{5} & \ \ldots \ & & & &\rstick[4]{\ket{0^{n-1}}}\\
\wave&&&&&&&&&&&& &
\\[0.75cm]
& & &  &\ \ldots\ & \ctrl{5}  &  & & & \ \ldots \ & \ctrl{5} & & &\\
&&&& \ \ldots \ && \ctrl{5} & & &\ \ldots \ & & \ctrl{5}&&\\
\lstick[5]{\ket{\psi_{2^t}}} && \swap{5}  & &\ \ldots\ & & & \swap{6} & & \ \ldots \ & & & &\\ 
&&& \swap{5}& \ \ldots \ &&&&   \swap{6} & \ \ldots \ &&&&\\
\wave&&&&&&&&&&&&&
\\
& &  &  &\ \ldots\ & \swap{5} &&&& \ \ldots \ & \swap{6}  &&&\\
&&&&\ \ldots \ && \swap{5} &&&\ \ldots \ &&  \swap{1} &&\\
\lstick[5]{\ket{0^n}} & & \targX{} & &\ \ldots\ & & & & &\ \ldots \ & & \targX{}&&\\ 
&&& \targX{} &\ \ldots \ &&& \targX{} &&\ \ldots \ &&&&\\
\wave&&&&&&&&&&&&&&
\\
&&&&\ \ldots \ &\targX{}&&&&\ \ldots \ &&&& \\
& &  &  &\ \ldots\ &  &  \targX{} & & & \ \ldots \ & \targX{} & & &
\end{quantikz}

\begin{quantikz}
 & \ctrl{1} & \\[7pt] & \swap{1} & \\[7pt] & \targX{} &  \end{quantikz} =
 \begin{quantikz}
   &  & \ctrl{1} &  &\\ 
  &\targ{}  & \ctrl{1} & \targ{} &\\
  & \ctrl{-1} & \targ{} & \ctrl{-1} &
\end{quantikz}

    \caption{Top: Depth-$2$ implementation of controlled $\mathsf{S}$. Bottom: Relation between controlled-$\mathsf{SWAP}$ and Toffoli.}
    \label{fig:shift}
\end{figure}
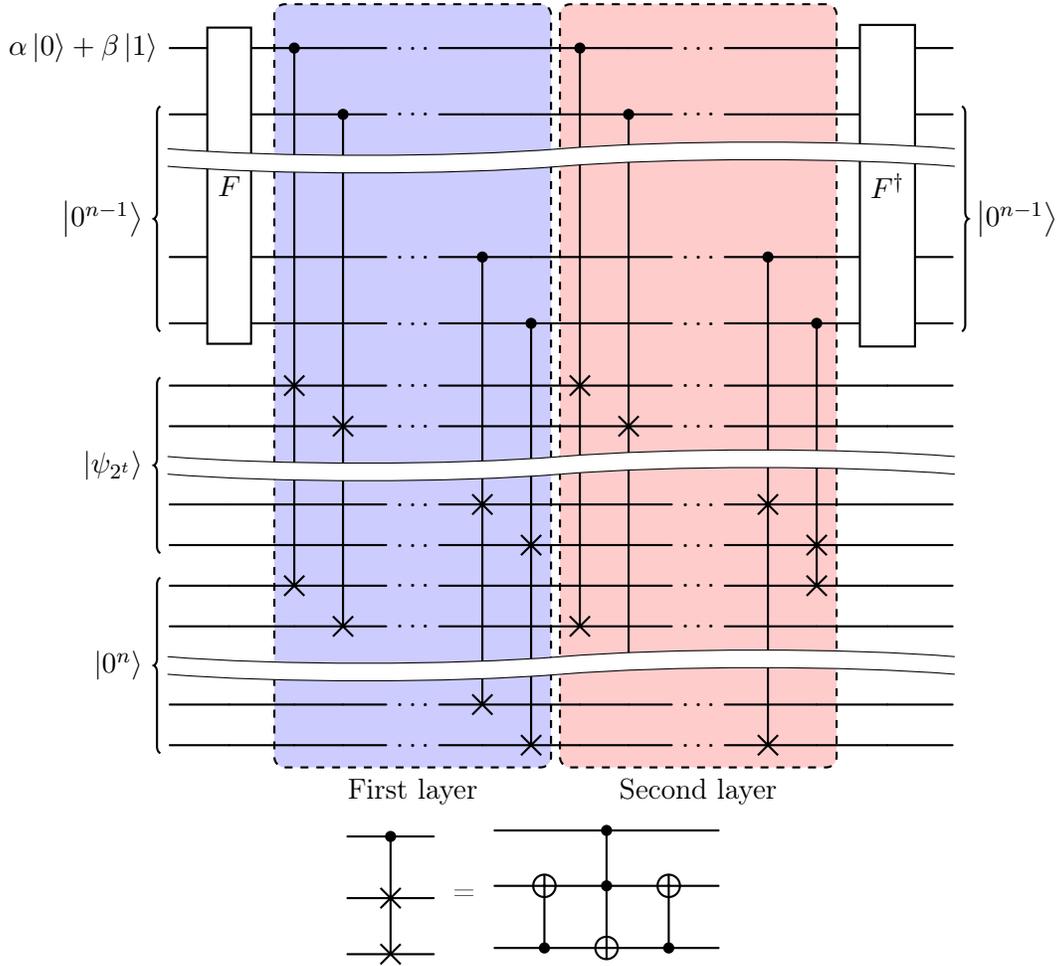

We remark that the Clifford depth of our method is also modest. Note that the only nonlocal gate we used is unbounded fanout. It is well known that this can be realized in a logarithmic-depth circuit consisting of $\mathsf{CX}$ gates~\cite{moore2001parallel}.

To summarize, insofar as $|\psi_{a2^t}\rangle$ is prepared offline for all $t\in [n]$ for some $a$ coprime to $2^n-1$, we can apply any rotation of $\frac{2\pi k}{2^n-1}$ for any $k\in [2^n-1]$ using three layers of Toffoli and logarithmically many layers of $\mathsf{CX}$ gates. The number of qubits used is $3n^2$. In order to ensure $m$-bit precision, it suffices to choose $n=m+1$.

\subsection{Resource Estimate}
\label{subsec:resource_estimate}

In Section~\ref{subsec:kickback}, we found that the $T$-depth  (or equivalently, the Toffoli depth) of our method is $3$. Here we quantify the number of non-Clifford gates used in our method and discuss a certain spacetime tradeoff we can make, depending on the number of non-Clifford gates one can apply in parallel. For this discussion, it will be convenient to phrase the result in terms of Toffoli gates. We also focus on the fixed-angle rotation for convenience. The resource estimate for the variable-angle rotation can be obtained by multiplying the non-Clifford gate count by a factor of $n$.

We also make what may seem at first like a strange assumption: that $n$ is at most $660$. The reason for making this assumption is the following. For any $n\leq 660$, a primitive polynomial that satisfies $|Q_f|=4$ exists~\cite{rajski2003primitive}. In fact, for many $n$, there are primitive polynomials with $|Q_f|=2$~\cite{brent2010great,arndt_primitive_trinomials_2003}. Let us discuss a few instructive examples. In Ref.~\cite[Appendix F]{beverland2022assessingrequirementsscalepractical}, the authors discuss two quantum computing applications that require precision of $\epsilon = 1.1\times 10^{-8}$ (for quantum dynamics simulation) and $\epsilon = 1.6\times 10^{-11}$ (for quantum chemistry simulation). This translates into using $n=27$ and $n=36$,  respectively. The following primitive polynomials will be suitable for this purpose~\cite{rajski2003primitive,arndt_primitive_trinomials_2003}:
\begin{equation}
\begin{aligned}
x^{27} + x^{20} + x^{13} + x^7+1 & \quad \text{ for } n=27,\\
    x^{36} + x^{11} + 1 & \quad \text{ for } n=36.
\end{aligned}
\end{equation}
Thus, for every $n\leq 660$, the controlled version of $\mathsf{CX}_{n-1 \to Q_f}$ uses at most $4$ Toffolis. We note that practically we will almost never need $n> 660$, because that corresponds to an astronomically high precision of $<4.78 \times 10^{-198}$. 

As for the controlled version of $\mathsf{S}$, we note that $\mathsf{S}$ can be implemented using $n-1$ $\mathsf{SWAP}$ gates in depth $2$ [Eq.~\eqref{eq:rev}]. Therefore, the controlled version of $\mathsf{S}$ uses at most $n-1$ Toffoli gates with a Toffoli depth of $2$. Combining this with the controlled version of $\mathsf{CX}_{n-1 \to Q_f}$, we arrive at a Toffoli count of at most $n+3$ (and sometimes $n+1$), Toffoli depth of $3$, and $n-1$ additional ancillary qubits needed for fanout. 

If we can apply at most $\kappa\geq 2$ Toffoli gates in parallel, it would make sense to only use $\kappa - 1$ additional ancillary qubits for the fanout. The overall Toffoli count would remain the same and the Toffoli depth would be $2 \lceil \frac{n-1}{\kappa-1} \rceil + \lceil \frac{4}{\kappa} \rceil$. Therefore, as the number of available non-Clifford gates increases, our method can be easily adapted to make it more time-efficient.

\subsection{Efficient Preparation of the Catalyst State}
\label{subsec:catalyst_prep}

Here we show that there is a polynomial-time quantum algorithm for preparing any $|\psi_k\rangle$. We actually describe two algorithms. The first algorithm is based on Shor's algorithm for the discrete logarithm problem~\cite{shor1994algorithms}. The second algorithm is based on the Frobenius endomorphism. The first algorithm is conceptually simpler, but it is less efficient. The complexity of both algorithms is dominated by the number of multiplications in $\mathrm{GF}(2^n)$. The first algorithm uses $\mathcal{O}(n)$ multiplications whereas the second algorithm uses the multiplication just three times; if we consider the cost \emph{per resource state}, the number of multiplications approaches one.

We remark that in both algorithms we use quantum phase estimation of $U_f$, which would yield an approximate eigenstate. In order to obtain an $\epsilon'$-approximation of $|\psi_k\rangle$ for some randomly chosen $k$, we would incur a cost polynomial in $\log (1/\epsilon')$ and $n$; see Section~\ref{subsubsec:discrete_log}. (Note $\epsilon'$ may be chosen to be different from the precision of the rotation angle, which we denoted as $\epsilon$ throughout this paper.)

In this section, it will be convenient to represent the elements of $\mathrm{GF}(2^n)$ as $\alpha^j, j\in [2^n-1]$. Recall that we chose $f(x)$ to be primitive, so this covers all the nonzero elements. So we write down $C_f^j v$ as $\alpha^{j+ j_0}$. We set $j_0=0$ for convenience, after which $|\psi_k\rangle$ becomes
\begin{equation}
    |\psi_k\rangle = \frac{1}{\sqrt{2^n-1}} \sum_{j=0}^{2^n-2} \tilde{\omega}^{-jk} |\alpha^{j}\rangle.
\end{equation}

\subsubsection{Algorithm Based on Discrete Logarithm}
\label{subsubsec:discrete_log}
Consider any nonzero binary vector, say $|10\ldots 0\rangle$, and apply quantum phase estimation for $U_f$. Since this is a Clifford, $U_f^{2^t}$ is a Clifford for all $t$. Moreover, they can be obtained by squaring. Thus the quantum phase estimation up to a precision $\epsilon'$ can be performed in gate cost polynomial in $\log (1/\epsilon')$ and $n$. Indeed, this entails applying $O(\log 1/\epsilon')$ layers of controlled $U_f^{2^t}$, each of which has $O(n^2)$ complexity. Because the overlap of the initial state with any $|\psi_k\rangle$ is $\frac{1}{2^n-1}$, the post-measurement state will be an $\epsilon'$-approximation of $|\psi_k\rangle$, where $k$ is random. 

Without loss of generality, suppose we obtained $|\psi_{k}\rangle$. We show that there is a polynomial-time algorithm that converts $|\psi_{k}\rangle$ to any other $|\psi_{k'}\rangle$. Note that 
\begin{equation}
    |\psi_{k'}\rangle = U_{k'-k}|\psi_k\rangle,
\end{equation}
where $U_m$ is defined as 
\begin{equation}
    U_m  |\alpha^{j}\rangle = \tilde{\omega}^{-jm} |\alpha^j\rangle.
\end{equation}

The unitary $U_m$ can be realized in two steps. The first step is to infer $j$ from $\alpha^j$, and the second step is to apply the phase proportional to $j$. The second step can be approximated well by the standard phase gradient method~\cite{kitaev2002classical}. So we focus on the first step, which is more nontrivial. 

Inferring $j$ from $\alpha^j$ is precisely the discrete logarithm problem. Famously, there is a polynomial-time quantum algorithm that can solve this problem~\cite{shor1994algorithms}. Therefore, there is a polynomial-time algorithm that implements $|\alpha^j\rangle |0\rangle \mapsto |\alpha^j v\rangle |j\rangle$. Knowing $j$, one can apply the requisite phase of $\tilde{\omega}^{-jm}$. One can then uncompute $j$, by solving this discrete logarithm problem again. This completes the implementation of $U_m$, which is clearly polynomial-time. The main bottleneck of the algorithm is the modular exponentiation, consisting of $n$ modular multiplications.

\subsubsection{Algorithm Based on Frobenius Endomorphism}
\label{subsubsec:frobenius}

Frobenius endomorphism is an endomorphism of $\mathrm{GF}(2^n)$ defined as $g(\alpha) \mapsto g(\alpha)^2$ for any $g(\alpha)\in GF(2^n)$. Because the field has characteristic $2$, this operation is linear. Moreover, it is invertible because $g(\alpha)^{2^n}=1$ for any $g\in GF(2^n)$. Thus this map is in fact an automorphism. Let $U_F$ be a unitary that implements this automorphism:
\begin{equation}
    U_F|\alpha^j\rangle = |\alpha^{2j}\rangle.
\end{equation}
This is a Clifford because $g(\alpha)\mapsto g(\alpha)^2$ is linear.

Applying $U_F$ to $|\psi_k\rangle$, we get
\begin{equation}
\begin{aligned}
    U_F|\psi_k\rangle &= \frac{1}{\sqrt{2^n-1}} \sum_{j=0}^{2^n-2} \tilde{\omega}^{-jk} |\alpha^{2j}\rangle \\
    &= \frac{1}{\sqrt{2^n-1}} \sum_{j=0}^{2^n-2} \tilde{\omega}^{-jk\cdot 2^{-1}} |\alpha^j\rangle \\
    &= |\psi_{k\cdot 2^{-1}}\rangle.
\end{aligned}
\label{eq:squaring}
\end{equation}
where $2^{-1}$ is the inverse of $2$ modulo $2^n-1$. Applying the inverse of $U_F$ on both sides, we get $U_F^{\dagger} |\psi_k\rangle = |\psi_{2k}\rangle$. By repeatedly applying $U_F^{\dagger}$ $t \in [n]$ times, we obtain a catalyst state $|\psi_{k\cdot 2^{t}}\rangle$. Because $U_F$ is Clifford, no extra non-Clifford gate is needed for this conversion. 

With Eq.~\eqref{eq:squaring} in mind, we can now describe the algorithm. Similar to the first step in Section~\ref{subsubsec:discrete_log}, we begin by applying quantum phase estimation for $U_f$ to some nonzero binary vector. The post-measurement state is $|\psi_k\rangle$, where $k$ is uniformly random. We continue to the next step if $k$ is coprime to $2^n-1$. Otherwise, repeat until this condition is met. The probability of success is
\begin{equation}
    \frac{\varphi(N)}{N} \geq \frac{1}{ e^{\gamma} \log \log N + \frac{3}{ \log \log N}},
\end{equation}
where $\varphi(N)$ is Euler's totient function, $\gamma$ is Euler's constant, and $N=2^n-1$~\cite{rosser1962approximate}. Thus on average we need to repeat this procedure at most $\mathcal{O}(\log n)$ times. 

Next, we copy $|\psi_k\rangle$, following the approach of Cleve and Watrous~\cite{cleve2000fast}. They observed that with a modular subtraction (or instead, addition) one can efficiently ``clone'' a quantum Fourier state. An analogous thing to do here is to use modular division (or instead, multiplication). Define $U_{\text{mul}}$ as
\begin{equation}
    U_{\text{mul}} (|\alpha^j\rangle|\alpha^{j'}\rangle) = |\alpha^j\rangle |\alpha^{j+j'}\rangle.
\end{equation}
Note that 
\begin{equation}
    U_{\text{mul}} (|\psi_0\rangle |\psi_k\rangle) = |\psi_{-k}\rangle |\psi_k\rangle.
\end{equation}
(Note $|\psi_0\rangle$ can be prepared with probability of $1-1/(2^n-1)$ by preparing the uniform superposition state and applying phase estimation for $U_f$.) Applying the same procedure on $|\psi_{-k}\rangle$, we obtain a copy of $|\psi_k\rangle$. Finally, one can uncompute $|\psi_{-k}\rangle$. This takes a total of three multiplications. If we want to make $m$ copies, we can reuse $|\psi_{-k}\rangle$ multiple times and then uncompute. The number of multiplications is then $m+2$.

Now, applying $(U_F^{\dagger})^t$ to each copy, we can convert $|\psi_k\rangle$ to $|\psi_{k 2^t}\rangle$. Since this conversion is purely Clifford, the dominant cost of the algorithm comes from the multiplication in $\mathrm{GF}(2^n)$. The cost of multiplication for each $|\psi_{k 2^t}\rangle$ approaches one. This way, we obtain $|\psi_{k2^t}\rangle$, $t\in [n]$, where $k$ is coprime to $2^n-1$. This is enough to apply the phase kickback procedure described in Section~\ref{subsec:kickback}.

\section{Discussion}
\label{sec:discussion}

In this paper, we show that an arbitrary single-qubit rotation can be approximated arbitrarily well by a circuit whose $T$-depth is $3$, using a catalyst state. We note that this leads to several nontrivial facts about quantum circuits. A good way to understand this implication is in terms of the complexity class $\mathsf{QNC}^0_f$~\cite{hoyer2005quantum,takahashi2016collapse}. This is a class of quantum circuits consisting of constant-depth quantum circuits (with arbitrary one- and two-qubit gates), accompanied by unbounded fanout. Because every rotation can be replaced by a constant $T$-depth circuit and a catalyst state, we can conclude that any circuit in $\mathsf{QNC}^0_f$ can be simulated by a constant $T$-depth circuit with the help of the catalyst states. These catalyst states can be viewed as quantum advice, whose size is at most polynomial in the input size~\cite{nishimura2004polynomial}. Thus we conclude that $\mathsf{QNC}^0_f/\mathsf{qpoly}$~\cite{watts2019exponential} admits a finite universal gate set, which is Clifford+$T$. Interestingly, $\mathsf{QNC}^0_f$ includes many interesting circuits, such as $\mathsf{OR}$ (or equivalently, Toffoli), addition, quantum Fourier transform, and even the quantum part of the factoring algorithm~\cite{hoyer2005quantum}. Thus all these circuits can be implemented in constant $T$-depth, with the help of a catalyst state. 

We leave the following questions for future work. First, note that for fixed rotation angle the size of the catalyst state and the $T$-count is $\mathcal{O}(\log (1/\epsilon))$. However, if the rotation angle is variable, these numbers change to $\mathcal{O}(\log^2(1/\epsilon))$. Can there be a constant $T$-depth catalytic circuit with $\mathcal{O}(\log (1/\epsilon))$ $T$-count? Secondly, we leave it as an open problem to show/disprove that constant $T$-depth rotation is possible using only clean/dirty ancillas. Thirdly, our work shows that many subroutines, e.g., multi-qubit Toffoli, addition, quantum Fourier transform can be implemented in constant $T$-depth with some catalyst state. However, plugging in our result directly to Ref.~\cite{hoyer2005quantum} will likely yield impractically large circuits. We leave it as an open problem to devise compact and efficient catalytic constant $T$-depth circuits for these subroutines. Lastly, we note that the circuits for multi-qubit Toffoli, addition, and quantum Fourier transform using our approach will be only approximately correct; the rotation angles used in Ref.~\cite{hoyer2005quantum} are not integer multiples of $2\pi / (2^n-1)$. We leave it as an open problem to devise catalytic constant $T$-depth circuits that implement these operations exactly.

\textbf{Note}: After this preprint appeared, several improvements were made. Gidney pointed out that there is an alternative construction that uses $T$-depth of $2$~\cite{gidney_x_1936285631359197210}. Notably, this construction uses $\mathcal{O}(n)$-size catalyst state to implement arbitrary rotation. Gidney also pointed out that in the setup in which different Pauli rotations are applied in sequence, the amortized $T$-depth can approach $1$. More recently, it was pointed out that using measurement-based uncomputation the $T$-depth can be lowered to $1$, without amortization~\cite{kim2025cliffordtcircuitcontrolledconstant}.

\textbf{Author contribution statement}: ChatGPT model O3 was used for literature search, for finding invertible boolean functions that have large periods. 

\section*{Acknowledgements}
I thank Michael Beverland, Earl Campbell, ChatGPT (OpenAI, model O3), Craig Gidney, Daniel Litinski, Vadym Kliuchnikov, Tuomas Laakkonen, Natalie Parham, and Henry Yuen for helpful discussions. I thank Aram Harrow for pointing out Theorem 13.5 in Ref.~\cite{kitaev2002classical}. The figures in this paper are drawn using Quantikz~\cite{kay2023tutorialquantikzpackage}.

\bibliographystyle{quantum}
\bibliography{ref_updated}

\end{document}